\documentclass[a4paper,11pt]{article}
\usepackage{pos}

\usepackage{graphicx}
\graphicspath{{images/}}
\usepackage{amsmath}

\title{$\mathcal CP$-violating anomalous interactions at Large Hadron Collider}

\author*[a]{Apurba Tiwari}
\author[a]{Sudhir Kumar Gupta}

\affiliation[a]{Aligarh Muslim University,\\
  Aligarh-202002, UP, India}

\emailAdd{atiwari@myamu.ac.in}
\emailAdd{sudhir.ph@amu.ac.in}

\abstract{ 
In this study, we explore the effects of $\mathcal CP$-violating anomalous interactions of the top-quark through the semileptonic 
decay modes of the top-quark arising due to pair-production of $t\bar{t}$ at the Large Hardon Collider. Predictions on the LHC sensitivities 
of the coupling strength to such $\mathcal CP$-violating interactions would be discussed for the 13 TeV LHC data and for the future 
hadron collider with 14 TeV energy.
}

\FullConference{%
  The 10th Annual\\
  Large Hadron Collider Physics Conference\\
  16-20 May 2022 
}

\begin{document}
\maketitle

\section{Introduction}

The Standard-Model (SM) is known as the most complete and rich model that is capable of explaining a large number of experimental 
data. Despite this, SM is incomplete as it is unable to explain some phenomena such as $\mathcal CP$-violation 
\cite{Nir:1999mg,Grossman:1997pa}, neutrino mass \cite{Bilenky:1999mf,Chao:2012mx}, existence of dark matter and dark energy 
\cite{Sahni:2004ai} etc. This indicates that there should be extensions to the SM to answer these open questions that cannot be 
addressed by SM. In the present article, we explore the possibility to search for $\mathcal CP$-violation in the process of top-quark 
pair production via proton-proton collision in the semi-leptonic decay modes of the top (anti-top)-quark. In particular, we find the 
$\mathcal CP$-violation sensitivity to anomalous top-quark couplings using T-odd triple product correlations 
\cite{Alioli:2017ces,Gupta:2009wu} constructed through the momenta of the end state particles for the already present data at LHC with 
$\sqrt{S}$ = 13 TeV and its luminosity intense variant HL-LHC with $\sqrt{S}$ = 14 TeV. $\mathcal CP$-violating anomalous top-quark 
interactions have been studied extensively in the existing literature 
\cite{Bernreuther:2010ny,Dawson:2013owa,Antipin:2008zx,Gupta:2009eq,Tiwari:2022nli,Tiwari:2022pug}.

We consider the model independent approach where the $\mathcal CP$-violation in the top-pair production vertex is parameterised by 
anomalous top-quark coupling. The SM Lagrangian could be modified through following effective Lagrangian where the modification in the 
top-pair production vertex occurs through the additional term in the Lagrangian \cite{Gupta:2009eq}:
\begin{eqnarray}
\label{intlag}
\mathcal L_{int}&=&
-i\frac{g_s}{2}\left(\frac{d_g}{\Lambda}\right)\bar{t}\sigma_{\mu\nu}\gamma_5\ G^{\mu\nu}\ t,
\end{eqnarray}

where $g_s$ is the strong coupling constant, $G^{\mu\nu}$ is the gluon field-strength tensor, $d_g$ is the interaction strength and 
$\Lambda$ is the $\mathcal CP$-violation energy scale. We consider the following T-odd correlations induced by the anomalous top-quark 
couplings in the process $pp \to t\bar{t} \to (bl^{+}\nu_l)(\bar{b}l^{-}\bar{\nu_l})$ \cite{Tiwari:2019kly}.
\begin{eqnarray}
\label{obs}
\nonumber
\mathcal C_1 &=&\epsilon(p_b,p_{\bar{b}},p_{l^+},p_{l^-}),\\
\nonumber
\mathcal C_2 &=& \tilde q \cdot (p_{l^+}-p_{l^-})~\epsilon(p_{l^+},p_{l^-},p_{b}+p_{\bar{b}},\tilde q),\\
\nonumber
\mathcal C_3 &=& \tilde q \cdot (p_{l^+}-p_{l^-})~\epsilon(p_b,p_{\bar{b}},p_{l^+}+p_{l^-},\tilde q),\\
\nonumber
\mathcal C_4 &=& \epsilon(P,p_b-p_{\bar{b}},p_{l^+},p_{l^-}),\\
\mathcal C_5 &=& \epsilon(p_b + p_{l^+},p_{\bar{b}} + p_{l^{-}},p_b+p_{\bar{b}},p_{l^+}-p_{l^-}),
\end{eqnarray}

where $\epsilon(a,b,c,d)$ is defined as $\epsilon(a,b,c,d) = \epsilon_{\mu \nu \alpha \beta} a^{\mu} b^{\nu} c^{\alpha} d^{\beta}$ with 
$\epsilon_{0123} = 1$, which is a completely anti-symmetric tensor and represents the Levi-Civita symbol of rank 4, $p_b~(p_{\bar{b}})$ 
represents the momenta of the $b~(\bar{b})$-quark and $p_{l^+}~(p_{l^-})$ represents the momenta of the lepton(anti-lepton) that emerges from 
$W^{+}~(W^{-})$ boson. P and $\tilde q$ are the sum and difference of the two initial proton beams which can be numerically defined as
\begin{eqnarray}
\label{P_q eqn}
\nonumber
P = p_b + p_{l^{+}} + p_{\bar{b}} + p_{l^{-}},~\tilde q = P_1 - P_2.\\
\end{eqnarray}

All observables defined in equation \ref{obs} are odd under $\mathcal CP$-transformation and take the form of triple product $\vec 
p_1.(\vec p_2 \times \vec p_3)$, where $p_i$ (i=1,2,3) represents the momentum vectors.

\section{Numerical Analysis}

We begin the simulation by incorporating the Lagrangian given in Eq. \ref{intlag} in the $\tt FeynRules$ \cite{Alloul:2013bka}, 
which then interfaced with the $\tt Madgraph5$ \cite{Frederix:2018nkq} for the event generation of $t\bar{t}$ pair via the process $pp 
\to t\bar{t}$ at LO. The events produced were then interfaced with $\tt pythia8$ \cite{Bierlich:2022pfr} for shower and Hadronisation. 
The values of the SM input parameters used in our study are given in Table \ref{SMinputs}, the value of renormalisation and 
factorisation scale is taken to be $M_Z$ and the parton distribution functions (PDF) considered was nn23lo1. We perform the analysis 
for LHC with $\sqrt{S}$ = 13 TeV for the integrated luminosity of 36.1 fb$^{-1}$ and 140 fb$^{-1}$ and High Luminosity LHC HL-LHC with 
$\sqrt{S}$ = 14 TeV for the projected luminosities of 0.3 ab$^{-1}$ and 3 ab$^{-1}$. The following cuts are implemented for generating 
the events:
\begin{eqnarray}
\label{cuts}
\nonumber
P_T(l^\pm) > 20~{\rm GeV},~ P_T(b,\bar{b}) > 25~{\rm GeV},~ \eta(b,\bar{b},l^\pm) < 2.5, \\
\Delta R (b\bar{b}) > 0.4,~ \Delta R (l^{+}l^{-}) > 0.2,~ \Delta R (bl) > 0.4,~ \not{E_T} > 30~{\rm GeV}.
\end{eqnarray}

\begin{table}[!ht]
\centering
\scalebox{0.9}{
\renewcommand{\arraystretch}{1.2}
\begin{tabular} {|c|c|}
\hline
SM parameter & Experimental value\\
\hline
$m_b (m_b)$ & 4.7 $\pm$ 0.06 GeV\\
$m_t(m_t)$ & 173.0 $\pm$ 0.4 GeV\\
$M_W$  & 80.387 $\pm$ 0.02 GeV \\
$\alpha_s^{\overline {MS}}(M_z)$ & 0.118 $\pm$ 0.001\\
 \hline
\end{tabular}}
\caption{Standard Model input parameter values \cite{ParticleDataGroup:2018ovx}.}
\label{SMinputs}
\end{table}

\begin{figure*}[h!]
\begin{tabular}{c c}
\hspace{-0.3cm}
\includegraphics[width=0.49\textwidth,height=5.0cm]{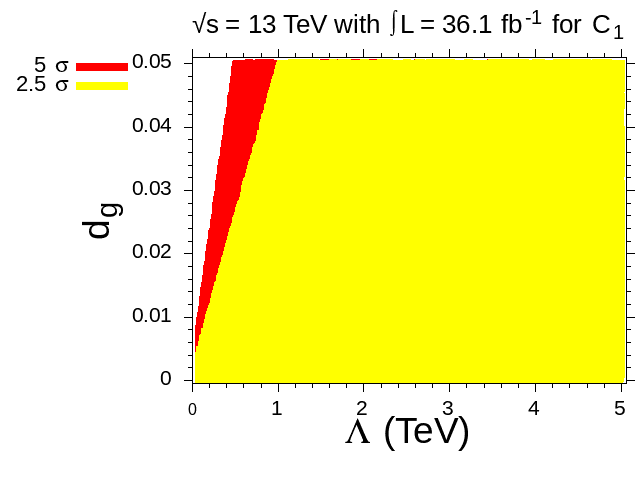}\hspace{-0.3cm}
&\includegraphics[width=0.49\textwidth,height=5.0cm]{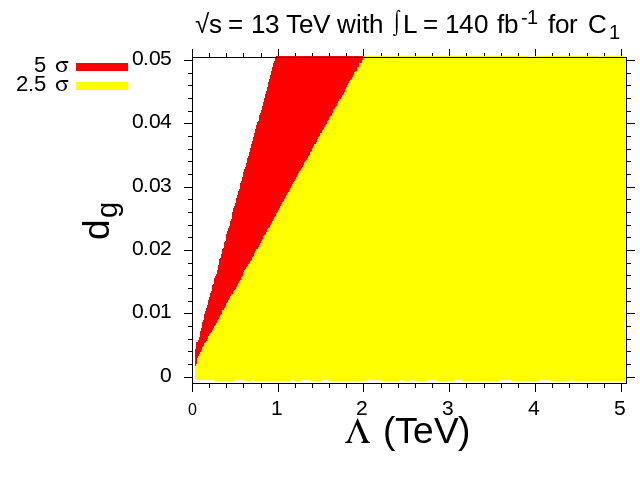}\hspace{-0.3cm}
\end{tabular}
\caption{The 2.5$\sigma$ (yellow) and 5$\sigma$ (red) regions in $d_g$-$\Lambda$ plane allowed by the production asymmetries in the 
semileptonic decay modes at LHC with $\sqrt {S}$ = 13 TeV for the integrated luminosities of 36.1 fb$^{-1}$ (left) and 140 fb$^{-1}$ (right) 
respectively.}
 \label{Plot13TeVC1}
\end{figure*}

\begin{figure*}[h!]
\begin{tabular}{c c}
\hspace{-0.3cm}
\includegraphics[width=0.49\textwidth,height=5.0cm]{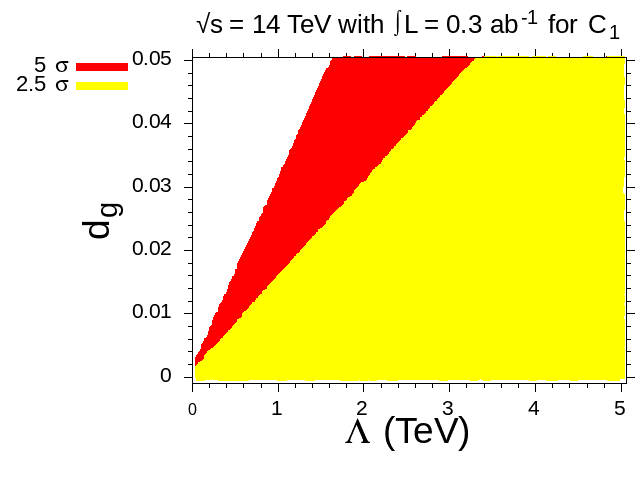}\hspace{-0.3cm}
&\includegraphics[width=0.49\textwidth,height=5.0cm]{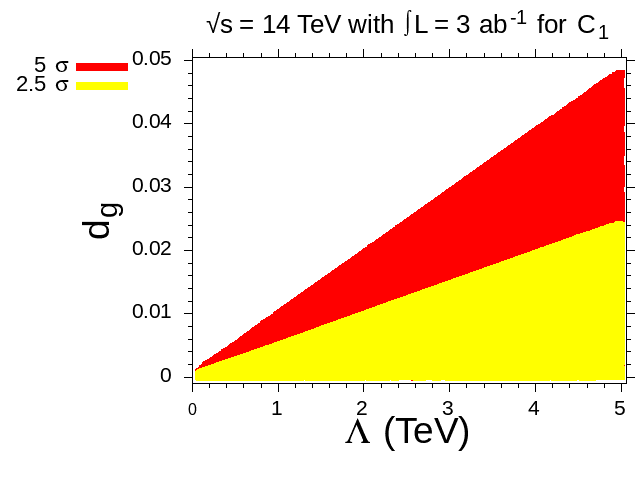}\hspace{-0.3cm}
\end{tabular}
 \caption{The 2.5$\sigma$ (yellow) and 5$\sigma$ (red) regions in $d_g$-$\Lambda$ plane allowed by the production asymmetries in the 
semileptonic decay modes at HL-LHC with $\sqrt {S}$ = 14 TeV for the integrated luminosities of 0.3 ab$^{-1}$ (left) and 3 ab$^{-1}$ (right)
respectively.}
 \label{Plot14TeVC1}
\end{figure*}

We use the method of finding asymmetries to study $\mathcal CP$-violation. A non-zero value of asymmetry would indicate the presence 
of $\mathcal CP$-violation. For this, we first produce events for the process concerned for different sets of values of the coupling 
constant and the $\mathcal CP$-violation scale $(d_g, \Lambda)$. The obtained results were then used for further analysis. In Figs. 
\ref{Plot13TeVC1} and \ref{Plot14TeVC1}, we show the 2.5$\sigma$ CL and 5$\sigma$ CL regions in $d_g$-$\Lambda$ plane allowed by the 
production asymmetries in the semileptonic decay modes at LHC with $\sqrt {S}$ = 13 TeV for the integrated luminosities of 36.1 
fb$^{-1}$ and 140 fb$^{-1}$ and at HL-LHC with $\sqrt {S}$ = 14 TeV for the integrated luminosities of 0.3 ab$^{-1}$ and 3 ab$^{-1}$. 
As we can see in the Figs. \ref{Plot13TeVC1} and \ref{Plot14TeVC1}, we have a wide set of ($d_g,~\Lambda$) values to achieve 5$\sigma$ 
sensitivity for $\mathcal CP$-violating anomalous top-quark coupling. The bounds on the coupling $\tilde d_g$ at LHC with $\sqrt{s}$ = 
13 TeV and HL-LHC with $\sqrt{S}$ = 14 TeV for the luminosities of 36.1 fb$^{-1}$ to 3 ab$^{-1}$ are given in Table 
\ref{sens_results}.

\section{Conclusions}

We have explored $\mathcal CP$-violating anomalous couplings in the top-pair production vertex through the process $pp \to t\bar{t} 
\to (bl^{+}\nu_l)(\bar{b}l^{-}\bar{\nu_l})$. In particular, we find constraints on $\mathcal CP$-violating anomalous top couplings by 
measurement of production asymmetry in the semileptonic detection modes. The sensitivity to $\mathcal CP$-violating coupling at 
2.5$\sigma$ CL and 5$\sigma$ CL for LHC with $\sqrt{S}$ = 13 TeV and HL-LHC with $\sqrt{S}$ = 14 TeV for luminosities of 36.1 
fb$^{-1}$, 140 fb$^{-1}$ and 0.3 ab$^{-1}$, 3 ab$^{-1}$, respectively are presented in Table \ref{sens_results}.

\begin{table}[h!]
\centering
\scalebox{0.9}{
\renewcommand{\arraystretch}{1.2}
\begin{tabular} {|c|c|c|c|}
\hline
$\sqrt{S}~(\rm{TeV})$ & $\int \mathcal L dt$ & \multicolumn{2}{|c|}{$\left|\frac{d_g}{\Lambda}\right|~(\rm{in~GeV}^{-1})$}\\
\hline
 &  & $\rm{at}~3\sigma~\rm{C.L.}$ & $\rm{at}~5\sigma~\rm{C.L.}$\\
  \hline
13 & 36.1 fb$^{-1}$ & 0.29 $\times 10^{-4}$ & 0.6 $\times 10^{-4}$\\
 & 140 fb$^{-1}$ & 0.52 $\times 10^{-5}$ & 0.2 $\times 10^{-4}$\\
\hline
14 (HL-LHC) & 0.3 ab$^{-1}$ & 0.39 $\times 10^{-5}$ & 0.6 $\times 10^{-5}$\\
 & 3.0 ab$^{-1}$ & 0.14 $\times 10^{-4}$ & 0.1 $\times 10^{-4}$\\
 \hline
\end{tabular}}
\caption{bounds on coupling $\tilde d_g~\left(\left|\frac{d_g}{\Lambda}\right|\right)$ at 3$\sigma$ CL and 5$\sigma$ CL in the context of 
top-pair production through the process $pp \to t\bar{t} \to (bl^{+}\nu_l)(\bar{b}l^{-}\bar{\nu_l})$ at LHC with $\sqrt{S}$ = 13 TeV and 
HL-LHC with $\sqrt{S}$ = 14 TeV for the luminosities of 36.1 fb$^{-1}$, 140 fb$^{-1}$ and 0.3 ab$^{-1}$, 3 ab$^{-1}$, respectively.}  
\label{sens_results}
\end{table}

\end{document}